\documentstyle[12pt]{article}
\normalsize

\newcounter{sxn}
\def\sx#1{\addtocounter{sxn}{1} \bigskip\medskip \goodbreak \noindent{\large\bf
\centerline{\thesxn.~~#1}} \nobreak \medskip}
\def\sxn#1{\sx{#1} }

\newcounter{axn}

\def\br{\begin{thebibliography}{abc}}
\def\er{\end{thebibliography}}

\date{}

\begin{document}
\bibliographystyle{unsrt}
\footskip 1.0cm
\thispagestyle{empty}

\setcounter{page}{0}
\begin{flushright}
UAHEP-933 \\
April 1993
\end{flushright}
%\vspace*{5mm}
\begin{center}{\LARGE DEFORMED WONG PARTICLES \\ }
\vspace*{6mm}
{\large   A. Stern and I. Yakushin\\}
\newcommand{\bc}{\begin{center}}
\newcommand{\ec}{\end{center}}
\vspace*{5mm}
 {\it Department of Physics, University of Alabama, \\
Tuscaloosa, AL 35487, USA.}\ec

\vspace*{5mm}

\normalsize
\centerline{\bf ABSTRACT}
By generalizing the Feynman proof of
the Lorentz force law, recently reported by Dyson,
we derive equations of motion for particles possessing internal
degrees of freedom $I^a$ which do not, in general, generate a finite
algebra.   We obtain consistency criteria for the fields which
interact with such particles.
It is argued that when a particle with internal $SU_q(2)$
degrees of freedom is coupled to $SU(2)$ gauge fields, $SU(2)$
gauge invariance is broken to $U(1)$.  We further claim that
when such an $SU_q(2)$ particle acts as a source for the field theory,
the second rank
antisymmetric field tensor, in general, cannot be globally defined.

\newpage
\newcommand{\be}{\begin{equation}}
\newcommand{\ee}{\end{equation}}
\baselineskip=24pt
\newcommand{\ba}{\begin{eqnarray}}
\newcommand{\ea}{\end{eqnarray}}
\newcommand{\no}{\nonumber}

\sxn{Introduction}

Recently there has been interest in constructing gauge theories based on
quantum algebras\cite{cas}.  Among the motivations for this activity
is the hope of introducing a new symmetry breaking mechanism in gauge
theories which could eventually be used to generate masses for vector
fields.  In the previous approaches, the
Lie algebras associated with gauge fields were deformed, with the
resulting field components having nontrivial commutation properties.
In this article, we shall
rather be interested in deforming the Lie algebras associated
with particles which can couple to gauge fields.  For us, the field
components are c-numbers at the classical level.
More generally,
we shall examine the dynamics of a particle possessing internal
degrees of freedom $I^a$ that do not generate a finite
algebra.  To obtain the dynamics,
we need only postulate the particle's Poisson brackets
and assume the existence of a Hamiltonian evolution.
The procedure is completely analogous to the Feynman proof of
the Lorentz force law, recently reported by Dyson\cite{dys}.
The proof has been
 generalized to the case of a particle interacting with gravity
and Yang-Mills fields in ref. \cite{tan}, and
a particle with external (or spin) $SU_q(2)$ degrees interacting with
a magnetic field in ref. \cite{son}.  Our results offer a further
generalization.  They can be applied
to the case of a particle with internal (or isospin) $SU_q(2)$
degrees of freedom interacting with an $SU(2)$ gauge field.  We
argue that when this happens, $SU(2)$
gauge invariance is broken to $U(1)$.
 We later obtain consistency criteria for fields having
such a particle as a source and claim that in this case second rank
antisymmetric field tensors cannot, in general, be globally defined.

The outline of this article is as follows:  We obtain the generalized
Lorentz force law in Section 2.  In Section 3, we
 apply it to the case where $I^a$
span a Lie algebra and in so doing recover the Wong particle
equations\cite{won}.  The case where $I^a$
defines an $SU_q(2)$ algebra, and
the breaking of $SU(2)$ gauge invariance is discussed in Section 4.
Finally, the consistency criteria for the associated
field equations is examined in Section 5.

\sxn{The Generalized Particle Equations}

We begin by examining a classical nonrelativistic particle of mass $m$.
The relativistic generalization will be considered later.
We denote the spatial coordinates of the particle
by $x^i$ and its velocity by $v^i$, $i=1,2,3$.
They are functions of some time parameter $t$, and
$v^{i}=\dot x^i$, where the dot denotes time differentiation.
For Poisson brackets involving $x$ and $v$, we postulate the following:
\ba
\{x^i,x^j\}&=&0   \;,        \\
\{x^i,v^j\}&=&{1\over m}\delta^{ij}\;.
\ea

Next, we introduce an internal degree of freedom which we
denote by $I^a=I^a(t)$, $a=1,...,D$,
and assume the general Poisson bracket relations
\ba
\{I^a,I^b\}&=&C^{ab}(I)\;,    \\
\{x^i,I^a\}&=&0     \;.
\ea
More generally, we assume that for arbitrary functions $A=A(x,I)$
and $B=B(x,I)$ we can write
$$
\{A,B\}= C^{ab}(I)\;\delta_a A \;\delta_b B
$$
in any local region of phase space, where
$\delta_a$ denotes a derivative with respect to $I^a$.
Here we shall make no special assumptions for the function
$C^{ab}(I)$, except
\be
C^{ab}=-C^{ba}\quad {\rm and} \quad
\delta_d C^{bc} C^{ad} + \delta_d C^{ca} C^{bd} +
\delta_d C^{ab} C^{cd} = 0  \;,
\ee
which follows from
antisymmetry of the Poisson bracket and the Jacobi identity.
Thus the $I$'s need not generate a finite algebra.

{}From the above Poisson bracket relations, along with the assumption that
$\dot v^i$ and $\dot I^a$ are functions of $x$, $v$, $I$ and $t$ only,
we can show that the equation of motion for the particle must be
of the form
\be
m \dot v^i = {\cal F}^{ij}(x,t,I)v^j +{\cal F}^{i0} (x,t,I)\;\;,
\ee
\be
\dot I^a =- {\cal A}^{i a}(x,t,I)v^i -{\cal A}^{0a} (x,t,I)\;\;,
\ee
where the ``fields" ${\cal F}^{\mu \nu}(x,t,I)
=-{\cal F}^{\nu\mu}(x,t,I)$ and ``potentials"
${\cal A}^{\mu a}(x,t,I)\; [\mu\nu,...=0,1,2,3 ] $
satisfy the following consistency conditions:
\be
{\cal D}^\lambda {\cal F}^{\mu \nu}  +
{\cal D}^\mu {\cal F}^{\nu \lambda}  +
{\cal D}^\nu {\cal F}^{\lambda \mu}  =0     \;,
\ee
\be
\delta_d  {\cal F}^{\mu \nu} C^{ad}=
{\cal D}^\mu {\cal A}^{\nu a} -{\cal D}^\nu {\cal A}^{\mu a}  \;,
\ee
\be
\delta_d C^{ab} {\cal A}^{\mu d} =
 \delta_d {\cal A}^{\mu b} C^{ad} - \delta_d {\cal A}^{\mu a} C^{bd} \;,
\ee
We define ${\cal D}^\mu$ by
$$
{\cal D}^\mu \equiv  \partial^\mu  - {\cal A}^{\mu d} \delta_d  \;\;,
$$
$\partial^j$ and $\partial^0$
denoting partial differentiation with respect to $x_j$
and $t$, respectively.

Following Feynman\cite{dys}, to prove these results we define
${\cal F}^{i j}$ and ${\cal A}^{i a}$ according to
\be
{\cal F}^{i j} =-  {\cal F}^{j i}
\equiv m^2 \{v^i ,v^j\} \quad
{\rm and}\quad {\cal A}^{i a} \equiv m \{v^i , I^a \} \;.
\ee
By applying the Jacobi identity involving $x^i$, $v^j$ and
$v^k$, we find that $x^i$ has zero Poisson bracket with
${\cal F}^{j k}$, while from
the Jacobi identity involving $x^i$, $v^j$ and
$I^a$, we find that $x^i$ has zero Poisson bracket with
${\cal A}^{j a}$.  It is for this reason that
${\cal F}^{i j}$ and ${\cal A}^{i a}$ can be functions of
$x$ and $I$ only.  The eq. (10) with index $\mu=i$
 follows simply from the Jacobi
identity involving $I^a$, $I^b$ and $v^i$.
Additional Jacobi identities give the following
consistency conditions on $C^{ab}$, ${\cal F}^{i j}$
and ${\cal A}^{i a}$:
\be
{1\over m}\delta_d  {\cal F}^{i j} C^{ad}=
\{v^j,{\cal A}^{i a} \}  -\{v^i,{\cal A}^{j a} \} \;,
\ee
\be
\{v^i,{\cal F}^{j k} \}  +
\{v^j,{\cal F}^{k i} \} +
\{v^k,{\cal F}^{i j} \} =0\;.
\ee

By taking the time derivative of the Poisson bracket relations,
and assuming the usual Leibniz rule,
we can deduce the general form for the equations of motion for
the system.  The time derivative of eqs. (2) and (4), leads to
\be
m^2 \{x^{i},\dot v^j \}  =  -{\cal F}^{i j}(x,t,I)  \quad{\rm and}
\quad m \{x^i,\dot I^a \} =- {\cal A}^{i a}(x,t,I)  \;,
\ee
respectively.  Then from the Poisson brackets (1), (2) and (4),
$\dot v^i$ and $\dot I^a$ must be of the form (6) and (7).
We can obtain conditions on the function
${\cal A}^{0a}$ by taking the time derivative of (3), leading to
$$
\{\dot I^a,I^b\} - \{\dot I^b,I^a\}  = \delta_d C^{ab} \dot I^d \;.
$$
After substituting (7) and using (10) with index $\mu=i$,
we find condition (10) with index $\mu=0$.

The remaining conditions are obtained by taking the time derivative of
eqs. (11).  By differentiating $ {\cal A}^{i a} = m\{v^i , I^a \} \; $
with respect to $t$ and using (6) and (7), we get
$$
 \partial^0 {\cal A}^{i a} +
\partial^j {\cal A}^{i a} v_j -\delta_d {\cal A}^{i a}
{\cal A}^{j d}v_j -\delta_d {\cal A}^{i a} {\cal A}^{0d} \qquad
$$
\be
\qquad = \delta_d {\cal F}^{i j} C^{da} v_j
+\delta_d {\cal F}^{i0} C^{da}
 -m\{v^i,{\cal A}^{j a}\}v_j -m\{v^i,{\cal A}^{0a}\}  \;.
\ee
We can equate terms linear in $v_j$ and terms independent
of $v_j $, leading to the two separate conditions.  One of them is
\be
 {\cal D}^i {\cal A}^{j a}=
\partial^i {\cal A}^{j a} -\delta_d {\cal A}^{j a} {\cal A}^{i d}
   =-m\{v^i,{\cal A}^{j a}\}  \;,
\ee
where we have used (12).  If we more generally assume that
$$
{\cal D}^i f(x,I)  =-m\{v^i,f(x,I)\}  \;,
$$
for any function $f(x,I)$, then the other condition is just (9)
with $(\mu \nu)=(i0)$.
Also for this, we have used the result that
$\{v^i,{\cal A}^{j a}\} $ and $\{v^i,{\cal A}^{0a}\} $
are independent of $v$, which follows
from the Jacobi identity involving $x$, $v$, ${\cal A}^{ia}$,
and $x$, $v$, ${\cal A}^{0a}$, respectively.
Upon using (16), the condition (12) reduces to (9)
with $(\mu \nu)=(ij)$.

By differentiating $ {\cal F}^{j i} =m^2 \{v^j , v^i \} \; $
with respect to $t$ and using (6) and (7), we get
$$
\partial^0 {\cal F}^{ji} +
\partial^k {\cal F}^{j i} v_k -\delta_d {\cal F}^{j i}
{\cal A}^{k d}v_k -\delta_d {\cal F}^{j i} {\cal A}^{0 d} \qquad
$$
\be
\qquad =  m\biggl(
\{{\cal F}^{j k},v^i\}v_k
-\{{\cal F}^{i k},v^j\}v_k
+\{{\cal F}^{j0},v^i\}
-\{{\cal F}^{i0},v^j\} \biggr) \;.
\ee
 Again we can equate terms linear in $v_i$ and terms independent
of $v_i $, leading to the two separate conditions: (8)
with $(\mu \nu \lambda)=(ij0)$ and
$$
{1\over m}{\cal D}^k {\cal F}^{j i}  =
 \{v^j,{\cal F}^{i k}\} -\{v^i,{\cal F}^{j k}\}  \;.
$$
Substituting the latter into eq. (13) then gives (8)
with $(\mu \nu \lambda)=(ijk)$.
We thereby have verified all equations (6-10).

In the above proof we have assumed the Leibniz rule for the time
derivative acting on a Poisson bracket; that is,
$$
{d\over{dt}}\{A, B \}=\{\dot A,B\}+\{A,\dot B\}\;.
$$
 This may not be valid in general\cite{tan}.  However,
it is true if the system admits a Hamiltonian $H$,
and the equations of motion can
be written as Hamilton's equations of
motion using $H$.  For our system, we can find a Hamiltonian.  It is,
\be
 H= {m\over {2}} v^i v^i + H_I (x, I)  \;.
\ee
We can then write the equations of motion (6) and (7) (along with
$\dot x^j  =v^j$) according to:
\be
\dot I^a  =\{I^a , H \} \;, \quad
\dot x^j   = \{x^j, H \}  \;\quad{\rm and} \quad
\dot v^j   = \{v^j, H \} +{{\partial v^j}\over{\partial t}} \;,
\ee
if we assume (11) and the following Poisson brackets for the interaction
Hamiltonian $H_I$:
\be
m\{v^j, H_I \}= {\cal F}^{j0}
 -m{{\partial v^j}\over{\partial t}}
\quad {\rm and} \quad
\{I^a, H_I \}= -{\cal A}^{0a} \;\;.
\ee
Here we are allowing for an explicit time dependence in the velocities
$v^i$.  Later we shall give an expression for $v^i$ and $H_I$
in terms of canonical phase space variables.

It is easy to generalize this system to the relativistic case.
For this we can keep all the Poisson brackets (1-4), as well
as (11) and (20), while we replace the Hamiltonian (18) by
\be
 H= m \sqrt{1+ v^i v^i} + H_I (x, I)  \;.
\ee
Now $\dot x^j \ne v^j$; Rather, we have
 $\dot x^j =  v^j / \sqrt { v^iv^i+1}  $.  This is obtained
from the Hamilton's equations of motion (19), along with
\be
 {d\over{dt}}\biggl({ {m\dot x^\mu} \over
 { \sqrt{- \dot x^\rho \dot x_\rho}}} \biggr)
 = {\cal F}^{\mu \nu}(x,I) \dot x_\nu \;\;,
\ee
\be
\dot I^a =- {\cal A}^{\mu a}(x,I)\dot x_\mu    \;\;,
\ee
where $x_\mu$ are components of a
four-vector with $x_0=t$ and we use the Minkowski metric tensor
$ [\eta_{\mu\nu}] = $diag$(-1,1,1,1)$.

As defined so far, the above
theory resembles a Kaluza-Klein theory because the ``fields"
${\cal F}^{\mu \nu}$ and ``potentials" ${\cal A}^{ a}_\mu$
are functions of internal coordinates $I$, as well as space-time
coordinates $x$.
To reduce the theory to one which is defined on four dimensional space-
time, it is necessary to make certain
assumptions on the fields, such as they
factorize into space-time dependent
 and internal space dependent
pieces.  Ans\"atze for the fields must be consistent with the
conditions (8-10).  We shall also require
that the ans\"atze do not put restrictions on the
particle degrees of freedom.

Now define ${\cal A}^a={\cal A}^{ a}(x,I)$
to be the one-form on
Minkowski space, with components ${\cal A}^{\mu a}$.
For ${\cal A}^{  a}$ we choose the following:
\be
{\cal A}^{ a}(x,I) = g \; C^{ab}(I) A_b(x)   \;\;,
\ee
where $g$ is a constant and $ A_b$ is a one-form on space-time.
Eq. (24) satisfies eq. (10) for all values of $x$ and $I$,
and for any $C^{ab}(I)$ fulfilling eq. (5).
Upon substituting the ansatz (24) into eq. (9), we get
\be
C^{ab} (I)\;\biggl(  {1\over g} \;
 \delta_b {\cal F}(x,I) - d A_b (x)
 -{g\over 2} \; \delta_b
C^{de}(I) A_d(x)\wedge A_e (x) \biggr) = 0 \;,
\ee
${\cal F}$ being the two-form on
Minkowski space, with components ${\cal F}^{\mu \nu}$.
$d$ and $\wedge$ denote exterior derivative on Minkowski space
and exterior product, respectively.
Ignoring $I$ independent terms, eq. (25) is solved by
\be
  {1\over g} \; {\cal F}(x,I) =   d A_a \; I^a +
  {g\over 2}\;  A_a\wedge A_b\; C^{ab} (I)  \;.
\ee
With ${\cal A}^a$ and ${\cal F}$ given in eqs. (24) and (26),
eq. (8) follows as an identity.
 [More generally, we may add an $I$ independent two-form
to (26), and from (8) that two-form is closed.]

Starting from
the ansatz (24) it is now easy to find an explicit form for the
velocities $v^i$ and the interaction Hamiltonian
$H_I$ appearing in eq. (18) [or (21)]
in terms of canonical coordinates
and momenta.  We denote the canonical momenta by $p^i$ and
 assume that it
has zero Poisson brackets with the internal variables.  Thus,
$
\{p^i,p^j\}=\{p^i,I^a\}=0  \; {\rm and} \;
\{x^i,p^j\}=\delta^{ij}\;.
$
In terms of $x^i,p^i$ and $I^a$, we can define $v^i$ and $H_I$ by:
\be
mv_i(x,p,I)=p_i - g A_a^i (x) I^a
 \quad {\rm and} \quad
H_I(x,I)= - g A_a^0 (x) I^a  \;.
\ee
{}From these definitions we then recover eqs. (11) and (20).
The resulting Hamiltonian (18) [or (21)] is identical to that of
a Wong particle\cite{won}.

\sxn{The Wong Equations}

Yang-Mills theory is recovered when $C^{ab}$ are linear functions
of $I$, the coefficients being the structure constants associated
with some Lie algebra $ G$.  That is,
\be
C^{ab}(I)={c^{ab}}_d \; I^d \;,
\ee
with eqs. (5) corresponding to
$
{c^{ab}}_d=-{c^{ba}}_d \quad {\rm and} \quad
{ c^{bc}}_d \;{c^{ad}}_e + {c^{ca}}_d \;{ c^{bd}}_e +
{c^{ab}}_d \; {c^{cd}}_e = 0  \;.
$
Then from eqs. (24) and (26),
${\cal A}^{ a}$ and ${\cal F}$ are also linear functions of $I$,
$A_b= A_b(x) $ corresponding to Yang-Mills connection
one-forms and $g$ being the coupling constant.
If we write
\be
{\cal F}(x,I) =g\; F_d(x) I^d  \;,
\ee
then we can identify $F_d=F_d(x)$ with the field strength two-form
for Yang-Mills theory,
\be
 F_d= d A_d +{g\over 2} {c^{be}}_d A_b \wedge A_e  \;,
\ee
from which it follows that (8) is the usual
 Bianchi identity for Yang-Mills fields,
$$
dF_a + g {c^{bd}}_a A_b \wedge F_d = 0 \;\;.
$$
Eqs. (22) and (23) correspond to the Wong equations of
motion for a particle in a nonabelian gauge field\cite{won}.

\sxn{ The Q-Deformed Wong Equations}

For general functions $C^{ab}(I)$ of $I$,
${\cal F}(x,I)$ does not factorize as in eq. (29).
In the case of quantum algebras, some components $C^{ab}(I)$ are linear
functions of $I^a$, while others are nonlinear in $I^a$.
We next consider the example of $SU_q (2)$.

The $SU_q (2)$ algebra is standardly realized for quantum operators
${\bf I}^+$, ${\bf I}^-$ and ${\bf I}^0 $ by the commutation relations:
\be
 [ {\bf I}^{0} , {\bf I}^{\pm} ] = \pm {\bf I}^{\pm} \;,
\qquad
 [ {\bf I}^{+} ,{\bf I}^{-} ] =   [ 2 {\bf I}^{0} ]_q    \equiv
{ {q^{2 {\bf I}^0} - q^{-2 {\bf I}^0}}\over {q - q^{-1}} } \;.
\ee
(These commutation relations reduce to the $SU(2)$ algebra relations
in the limit $q \rightarrow 1$.)
To obtain the corresponding classical system, let us replace
the quantum operators
${\bf I}^+,\;{\bf I}^-$ and ${\bf I}^0 $ by classically commuting
 variables which we denote by
 $I^1 +iI^2$, $I^1 -iI^2$ and $I^3$, respectively.  We also
replace the commutation relations (31) by $i$ times Poisson brackets.
The result is:
\be
\{I^2,I^3\}=I^1 \;,\quad
\{I^3,I^1\}=I^2 \quad {\rm and }\quad
\{I^1,I^2\}={1\over 2}  [ 2I^3 ]_q    \;.
\ee
{}From these relations, we identify the functions $C^{ab}$ according to:
$   C^{23}=I^1\;\;, C^{31}=I^2 \;\; {\rm and }\;\;
C^{12}={1\over 2}  [ 2I^3 ]_q  \;.  $
Then we can write
\ba
{\cal A}^{ a}(x,I)& =&g\; \epsilon^{abc}A_b(x)\biggl( I^c
 + \delta^{c3}h(I^3) \biggr)\;,   \\
{\cal F}(x,I) & =& g\;F_a(x) I^a
 + g^2\;  A_1\wedge A_2 \; h(I^3)
 \;\;   \;,
\ea
$$
h(I^3) = {1\over 2}  [ 2I^3 ]_q -I^3 \;\;,
$$
where $F_a$ is the $SU(2)$ Yang-Mills two-form eq. (30) with
$ {c^{ab}}_d= \epsilon^{abd}$.
{}From eq. (8) we again get identities involving the fields.

Although we have not deformed the Wong
particle Hamiltonian (18) [or (21)],
we have deformed the Poisson brackets (32)
from the $SU(2) $ case, and consequently also
the equations of motion from the $SU(2) $ case.
We thereby obtain ``q-deformed" or $SU_q(2)$
Wong equations (22) and (23), with ${\cal A}^{ a}$ and ${\cal F}$
defined in (33) and (34).  From them we can show that
\be
(I^1)^2+ (I^2)^2+{1\over{2\ln q}}\biggl(
{ {q^{2I^3} + q^{-2I^3}}\over {q - q^{-1} } } -
{1\over{\ln q}}\biggl)
\ee
is a constant of the motion.  The term
${1\over{\ln q}}$ was subtracted in parenthesis so that (35)
converges in the limit of $q\rightarrow 1$.  In that
limit, it just becomes the classical analogue of the
quadratic Casimir operator for $SU(2)$.  It is not hard to show
that (35) has zero Poisson bracket with $I^a$, and consequently
 all phase space variables, for any value of $q$.  Hence, it is
the classical analogue of a Casimir operator for $SU_q(2)$.

For $SU_q(2)$,
${\cal F}(x,I)$ given in eq. (34) is invariant under infinitesimal
$U(1)$ gauge transformations.  Under such
transformations, $A_a$ and $I^a$ undergo the infinitesimal changes:
\be
\delta A_a = \delta_{a3} \;
 d\Lambda_3 + g \;\epsilon_{ab3} A_b \Lambda_3 \quad ,\quad
\delta I^a =  g\; \epsilon_{ab3} I^b \Lambda_3   \;,
\ee
where $\Lambda_3$ is an infinitesimal function of the particle
space-time coordinates.  It follows that the $SU_q(2)$ Wong
equation (22) is invariant under $U(1)$ gauge transformations.  The same
is true for the equation of motion (23) using (33).

If we were to interpret $A_a$ as connection one-forms for
$SU(2)$ Yang-Mills theory,  then
eq. (36) corresponds to a $U(1)$ subgroup of $SU(2)$ Yang-Mills
transformations.  Under the full set of
$SU(2)$ Yang-Mills transformations
 an infinitesimal change in $A_a$ is given by
\be
\delta A_a =  d\Lambda_a + g \epsilon_{abc} A_b \Lambda_c \; ,
\ee
where $\Lambda_a$ are infinitesimal functions of the
space-time coordinates.
Is there a compensating transformation on $I^a$ such
that the $SU_q(2)$ Wong equations are invariant for $\Lambda_1,
\Lambda_2 \ne 0$?  We now argue that the answer is no.  In order that
${\cal F}(x,I)$ given in eq. (34) be invariant under (37),
$I^a$ must undergo a change $\delta I^a$, which
satisfies the following equation:
$$
F_b\;(\delta I^b + g \epsilon_{abc} I^a \Lambda_c) \;
+\;g\epsilon_{ab3}A_a \wedge \biggl(
A_b \;f(I^3) \delta I^3 +(d\Lambda_b +g \epsilon_{bcd} A_c \Lambda_d )
\;h(I^3)\biggr)\; =0\;,
$$
where $f(I^3) = {1\over 2}{{dh}\over {dI^3}}$.  But there exists
no solution for $\delta I^a$ for arbitrary $A_b$, and arbitrary
$\Lambda_1, \Lambda_2$.  To see this, consider gauge
transformations about the first axis, $\Lambda_a=
\delta_{a1}\Lambda_1$ and set
$F_2 =F_3=A_1=0$.  Then for arbitrary $A_2$,
the above condition states that
$$
A_3 \delta I^1 = d\Lambda_1 h(I^3) \;\;.
$$
But $A_3$ and $ d\Lambda_1$ are independent (closed) one-forms on
Minkowski space.  Hence, the condition cannot be satisfied.  The same
conclusion is reached when we consider
transformations about the second axis, $\Lambda_a=
\delta_{a2}\Lambda_2$ and set $F_1 =F_3=A_2=0$.
{\it We therefore conclude that
in the presence of $SU_q(2)$ Wong particles,
 the $SU(2)$ gauge invariance of
$SU(2)$ Yang-Mills theory is broken to $U(1)$.}

\sxn{The Field Equations}

In the above generalization of the Feynman proof of the Lorentz
force equation,
the dynamics of particles has been fully specified
 [eqs. (22) and (23)].  The same cannot be said about
the dynamics of fields.  Eqs. (8-10) are insufficient for
determining the field dynamics.  This, of course, was
also the case for electromagnetism.
The Gauss' law and Ampere's law actually
did not follow in the proof of
Feynman\cite{dys} and had to be postulated by hand.
Not surprisingly then, we too must postulate additional field equations
in order to fully specify the dynamics.
In these equations, particles now act as sources for the fields.
These equations are not completely arbitrary.  They must satisfy
certain consistency criteria
when the particle sources possess internal degrees of freedom $I^a$.
To see this, we introduce the space-time dependent quantities
$J^a_\mu(y)$ and $\Sigma^{ab}_\mu(y)=-\Sigma^{ba}_\mu(y)$
and define them as follows:
\ba
J^a_\mu(y)&=& \int dt\; \delta^4(y-x(t))\;I^a(t)\;
\dot x_\mu(t)\;, \\
 \Sigma^{ab}_\mu(y)
&=& \int dt\; \delta^4(y-x(t))\;C^{ab} [I(t) ]  \;\dot x_\mu(t)    \;.
 \ea
Here $x_\mu(t)$ and $I^a(t)$ are the space-time coordinates and
internal coordinates, respectively, of the source particle
and the integration is over the particle world line.
Now by multiplying the particle equation (23) by $ \delta^4(y-x(t))$
and integrating over $t$, we obtain the following relations:
\be
{{\partial }\over{\partial y_\mu}}J^a_\mu(y) + g A^\mu_b(y)
 \Sigma^{ab}_\mu(y) =0\;,
\ee
where we have assumed (24).  Now if we equate
$J^a_\mu(y)$ and $\Sigma^{ab}_\mu(y)$ to
some functionals of $A_\mu^a(y)$
(and possibly other fields) along with
their derivatives, then eqs. (38) and (39)
represent field equations for the system in the
presence of a source.  Furthermore, eq. (40) then corresponds
to a set of consistency criteria which the fields must satisfy.

For the case where the source is a Wong particle, the ``charges" $I^a$
and $C^{ab} [I]$ appearing in (38) and (39)
are linearly related [eq. (28)].  Consequently, so are
$\Sigma^{ab}_\mu$ and $ J^d_\mu$:
$ \Sigma^{ab}_\mu ={c^{ab}}_d J^d_\mu\;.$
Here, eq. (40) can be written
\be
(D^\nu J_\nu)^a=0\;,
\ee
$D^\nu \; \biggl( [ {D^\nu} { ]^{a}}_b=\partial^\nu\;\delta^a_b+
g{c^{ad}}_bA^\nu_d \biggr) $ denoting the covariant derivative.
This condition is just the statement that the Yang-Mills current
is covariantly conserved.  It is identically satisfied
with the usual choice of Yang-Mills field equations,
\be
(D^\nu F_{\nu\mu})^a =J^a_\mu  \;,
\ee
since upon substituting into
the condition (41), we get $ D^\mu(D^\nu F_{\nu\mu}) \equiv 0  \;.$

For the case where the source is a q-deformed Wong particle,
the ``charges" $C^{ab} [I]$ and $I^a$ are not all linearly
related, and no simple relation
exists between all the $\Sigma^{ab}_\mu$ and $ J^d_\mu$.
For $SU_q(2)$ particle sources, there are six quantities
 $\Sigma^{ab}_\mu$ and $J^a_\mu$, but from eqs. (32) we can make the
identifications:  $\Sigma^{23}_\mu=J^1_\mu$ and
 $\Sigma^{31}_\mu=J^2_\mu$.
Four independent quantities remain which we denote by
 $J^{(q)a}_\mu$ and $\Delta_\mu$.  We define them by:
\be
J^{(q)1}_\mu=J^1_\mu \;,  \quad
J^{(q)2}_\mu=J^2_\mu \;, \quad
J^{(q)3}_\mu=\Sigma^{12}_\mu
\quad   {\rm and} \quad
\Delta_\mu=\Sigma^{12}_\mu-J^3_\mu \;.
\ee
{}From eqs. (38) and (39) we thus have four field equations.
The conditions (40) can be written:
$$
(D^\mu J^{(q)}_\mu)^a= \partial^\mu \Delta_\mu\;\delta^{a3}  \;,
$$
where $D^\mu$ is the covariant derivative for $SU(2)$.
Using eqs. (38) and (39), we then have
\be
(D^\mu J^{(q)}_\mu)^a(y)= \delta^{a3}
\;\int dt\; \delta^4(y-x(t))\;{d\over{dt}} h(I^3(t))        \;.
\ee
In the limit $q\rightarrow 1$, $\Delta_\mu$ vanishes and
$J^{(q)a}_\mu\rightarrow J^a_\mu $.  Then eqs. (44)
 reduce to (41) and are solved by eq. (42).  We thereby recover
the field equations for $SU(2)$ gauge theory in that limit.

If we interpret
$J^{(q)a}_\mu$ as the current associated with a q-deformed
Wong particle, then eq. (44) shows that it is not covariantly
conserved for arbitrary $q$ and $I^a$.
[Here ``covariantly" means with regard to $SU(2)$ transformations.]
The same conclusion is reached upon taking
$J^{a}_\mu \;$ to be the current associated with q-deformed
particles.  In terms of this current, the condition (40) becomes:
$$\;(D^\mu J_\mu)^a= -g\;\epsilon^{ab3} A^\mu_b\Delta_\mu \;. $$

Let us examine the case where ${{dh}\over{dt}}$ is nonvanishing
only for a finite segment ${\bf L}$ (begining at a time $t=t_1$ and
ending at a time $t=t_2$) of an $SU_q(2)$ particle's world line.
Now if in analogy to Yang-Mills theory, we were then to write
\be
(D^\nu F_{\nu\mu})^a =J^{(q)a}_\mu \; \;,
\ee
we speculate that the fields components
$F^a_{\mu\nu}(y)$ may not all be globally defined on ${\bf M}
\setminus {\bf L},\;\;{\bf M}$ denoting Minkowski space.
To argue this point it is usefull to make the simplifying assumption that
all fields and potentials in the 1 and 2 directions of internal space
are zero.
This could be done consistently if we were allowed to set $I^1 =I^2 = 0$,
for all $t$, and hence $J^{(q)1}_\mu =J^{(q)2}_\mu =0$.
But with $I^1=I^2=0$ and $I^3$ changing, (35) cannot remain
a constant of the motion.  As an alternative, let us instead
imagine that the vector $I$ precesses rapidly about the third axis.
That is, $I^1$ and $I^2$ are oscillating rapidly (compared with $I^3$).
In other words, $I^1$ and $I^2$ are ``fast" variables, and $I^3$
is a ``slow" variable.   If we then time average over
the ``fast" variables we obtain the desired simplification;  Namely,
$J^{(q)1}_\mu =J^{(q)2}_\mu =0$.  The resulting time averaged fields
and potentials can then be made to point in the third direction in the
internal space.
Now define $*J^{(q)3}$ to be a 3-form whose components are dual to
$J^{(q)3}_\mu$.  From eq. (44), $d*J^{(q)3}$ is proportional to
$\int dt\; \delta^4(y-x(t))\;{d\over{dt}} h(I^3(t))\;.  $
If $S^3$ is a 3-sphere whose enclosing four dimensional
volume $V^4$ contains ${\bf L}$, then from Stoke's theorem
\be
\int_{S^3} *J^{(q)3}  \;=\;  \int_{V^4} d*J^{(q)3}  \;\;\; \propto \;\;\;
\int_{\bf L} dt\; \;{d\over{dt}} h(I^3(t))\;
= h(I^3(t_2))-  h(I^3(t_1)) \;.
\ee
It follows that if $ h(I^3) $ undergoes a nonzero change along ${\bf L}$,
then $*J^{(q)3}$ is a closed but not exact 3-form on $S^3$.  Thus
if we write $*J^{(q)3}=d*F^3$, the two-form $*F^3$ (and hence the
antisymmetric field tensor
 $F^3_{\mu\nu}$) cannot be defined everywhere on $S^3$.
 [More generally, if we define $*\Delta$ to be the 3-form whose
components are dual to $\Delta_\mu$
and if $ h(I^3) $ undergoes a nonzero change along ${\bf L}$,
then $* \Delta$ is closed but not exact on $S^3$.  Note that
for this to be valid no time averaging of the fields is necessary.]

The above is an adaptation of a result
found long ago by Rasetti and Regge\cite{reg}
in the context of superfluid helium.  There the
analogue of $F^3_{\mu\nu}$ was the antisymmetric potential
used to describe phonon excitations, while the source was interpreted
as corresponding to the injection of helium atoms in the superfluid.

{\bf Acknowledgements}

We are grateful to M. Lukin for discussions.
We have been supported during the course of this work
by the Department of Energy, USA under
contract number DE-FG-05-84ER-40141.

\enddocument